\documentclass[twocolumn]{aastex61}
\pdfoutput=1 
\usepackage{amsmath,amstext}
\usepackage[T1]{fontenc}
\usepackage{apjfonts} 
\usepackage[figure,figure*]{hypcap}
\usepackage{mypack}
\usepackage{graphicx}
\usepackage{verbatim}

\shorttitle{Masses of two binary pulsars discovered in the PALFA survey}
\shortauthors{Zhu et al.}

\begin{document}

\title{Mass measurements for two binary pulsars discovered in the PALFA survey}

\author{W.~W.~Zhu}
\affiliation{Max-Planck-Institut f\"{u}r Radioastronomie, Auf dem H\"{u}gel 69, 53131 Bonn, Germany}
\affiliation{CAS Key Laboratory of FAST, NAOC, Chinese Academy of Sciences, Beijing 100101, China;}

\author{P.~C.~C.~Freire}
\affiliation{Max-Planck-Institut f\"{u}r Radioastronomie, Auf dem H\"{u}gel 69, 53131 Bonn, Germany}

\author{B.~Knispel}
\affiliation{Max-Planck-Institut f\"{u}r  Gravitationsphysik,  D-30167  Hannover, Germany}
\affiliation{Leibniz Universit\"{a}t Hannover, D-30167 Hannover, Germany}

\author{B.~Allen}
\affiliation{Max-Planck-Institut f\"{u}r  Gravitationsphysik,  D-30167  Hannover, Germany}
\affiliation{Leibniz Universit\"{a}t Hannover, D-30167 Hannover, Germany}
\affiliation{Physics Dept., Univ. of Wisconsin - Milwaukee, Milwaukee WI 53211, USA}

\author{B.~W.~Stappers}
\affiliation{Jodrell Bank Centre for Astrophys., School of Phys. and Astr.,  Univ. of Manchester, Manch., M13 9PL, UK}

\author{A.~G.~Lyne}
\affiliation{Jodrell Bank Centre for Astrophys., School of Phys. and Astr.,  Univ. of Manchester, Manch., M13 9PL, UK}



\author{S.~Chatterjee}
\affiliation{Cornell Center for Astrophysics and Planetary Science and Dept. of Astronomy, Cornell Univ., Ithaca, NY 14853, USA}

\author{J.~M.~Cordes}
\affiliation{Cornell Center for Astrophysics and Planetary Science and Dept. of Astronomy, Cornell Univ., Ithaca, NY 14853, USA}

\author{F.~Crawford}
\affiliation{Dept. of Physics and Astronomy, Franklin and Marshall College, Lancaster, PA 17604-3003, USA}


\author{J.~S.~Deneva}
\affiliation{George Mason University, resident at the Naval Research Laboratory, 4555 Overlook Ave. SW, Washington, DC 20375, USA}

\author{R.~D.~Ferdman}
\affiliation{Faculty of Science, University of East Anglia, Norwich Research Park, Norwich NR4 7TJ, United Kingdom}

\author{J.~W.~T.~Hessels}
\affiliation{ASTRON, the Netherlands Institute for Radio Astronomy, Postbus 2, 7990 AA, Dwingeloo, The Netherlands}
\affiliation{Anton Pannekoek Institute for Astronomy, Univ. of Amsterdam, Science Park 904, 1098, XH Amsterdam, The Netherlands}

\author{V.~M.~Kaspi}
\affiliation{Dept. of Physics, McGill Univ., Montreal, QC H3A 2T8, Canada}

\author{P.~Lazarus}
\affiliation{Max-Planck-Institut f\"{u}r Radioastronomie, Auf dem H\"{u}gel 69, 53131 Bonn, Germany}

\author{R.~Lynch}
\affiliation{NRAO, Charlottesville, VA 22903, USA}


\author{S.~M.~Ransom}
\affiliation{NRAO, Charlottesville, VA 22903, USA}


\author{K.~Stovall}
\affiliation{National Radio Astronomy Observatory, 1003 Lopezville Road, Socorro, NM, 87801, USA}

\author{J.~Y.~Donner}
\affiliation{Max-Planck-Institut f\"{u}r Radioastronomie, Auf dem H\"{u}gel 69, 53131 Bonn, Germany}
\affiliation{Fakult\"{a}t f\"{u}r Physik, Universit\"{a}t Bielefeld, Postfach 100131, 33501 Bielefeld, Germany}

\correspondingauthor{Paulo C. C. Freire}
\email{pfreire@mpifr-bonn.mpg.de}

\begin{abstract}
In this paper, we present the results of timing observations of PSRs~J1949+3106 and
J1950+2414, two binary millisecond pulsars discovered in data from the Arecibo
ALFA pulsar survey (PALFA). The timing parameters include precise measurements
of the proper motions of both pulsars, which show that PSR~J1949+3106 has a transversal
motion very similar to that of an object in the local standard of rest. The timing also
includes measurements of the Shapiro delay and the rate of advance of periastron for both systems. Assuming
general relativity, these allow estimates of the masses of the components of the two systems;
for PSR~J1949+3106, the pulsar mass is $M_p \, = \, 1.34^{+0.17}_{-0.15} \, M_{\odot}$
and the companion mass $M_c \, = \, 0.81^{+0.06}_{-0.05}\, M_{\odot}$; 
for PSR~J1950+2414 $M_p \, = \, 1.496 \, \pm \, 0.023\, M_{\odot}$ and
$M_c \, = \, 0.280^{+0.005}_{-0.004}\, M_{\odot}$ (all values 68.3 \% confidence limits).
We use these masses and proper motions to investigate the evolutionary history of both systems:
PSR~J1949+3106 is likely the product of a low-kick supernova;
PSR~J1950+2414 is a member of a new class of eccentric millisecond pulsar binaries with an unknown formation mechanism. We discuss the proposed hypotheses for the formations of these systems
in light of our new mass measurements.
\end{abstract}

\keywords{(stars:) pulsars: general --- (stars:) pulsars: individual: PSR J1949+3106--- (stars:) pulsars: individual: PSR J1950+2414  --- stars: neutron --- binaries: general}

\section{Introduction}
\label{sec:introduction}

\subsection{The PALFA pulsar survey}

The PALFA survey \citep{2006ApJ...637..446C,2015ApJ...812...81L}, currently being carried out
with the Arecibo Observatory, has thus far discovered 189 pulsars\footnote{\url{http://www.naic.edu/~palfa/newpulsars/}}.
The high spectral and time
resolution of the data are optimized for the discovery of millisecond pulsars (MSPs,
defined here as recycled pulsars with a spin period $P < 25$ ms) at high values of dispersion measure
(DM), enabling searches for MSPs to great distances into the Galactic plane.
This approach is now well demonstrated, with the discovery of 30
new MSPs, most with large DMs compared to the previous population and large distances
\citep{2008Sci...320.1309C,2010Sci...329.1305K,2011ApJ...732L...1K,2012ApJ...757...90C,2012ApJ...757...89D,2013ApJ...773...91A,2015ApJ...800..123S,2015ApJ...806..140K,2016ApJ...833..192S}.
PALFA survey uses the {\tt PRESTO} software package \citep{2011ascl.soft07017R} for pulsar searches.

The instantaneous sensitivity of the Arecibo 305-m telescope means that the survey can achieve considerable
depth with relatively short pointings (about 4.5 minutes); this makes it sensitive to
highly accelerated systems. It is partly for this reason that the survey has already
discovered three new relativistic pulsar-neutron star (PSR - NS) systems:
PSRs J1906+0746 \citep{2006ApJ...640..428L}, a system where the pulsar we detect is the
second-formed  NS, J1913+1102 \citep{2016ApJ...831..150L}, a system with a likely mass
asymmetry and J1946+2052 \citep{2018ApJ...854L..22S}, the most compact
PSR - NS system known. This survey has also discovered the
first repeating fast radio burst~\citep{2014ApJ...790..101S,2016Natur.531..202S}.

\subsection{The pulsars}

PSR~J1949+3106 is one of the two new MSPs announced by \cite{2012ApJ...757...89D}. It has a spin period
of 13.1 ms, and it is in a binary system with an orbital period of 1.95 d. The projected
semi-major axis of its orbit ($x \, = \,$7.29 light seconds, lt-s)
implies that it has a massive companion; the small orbital eccentricity ($e \, = \, 0.000043$)
implies that this massive companion is a white dwarf star (WD). In the discovery paper, \citet{2012ApJ...757...89D} also measured the Shapiro delay in this system and determined
the masses of the pulsar and its companion:  $m_p \, = \, 1.47^{+0.43}_{-0.31}\, M_{\odot}$
and $m_c \, = \, 0.85^{+0.14}_{-0.11}\, M_{\odot}$. Although the Shapiro delay had been
detected with high significance, the uncertainties of the published masses
are too large to be astrophysically useful.

PSR~J1950+2414 was found by the Einstein@Home pipeline  \citep{2013ApJ...773...91A};
its discovery and early timing results were described by \cite{2015ApJ...806..140K}.
It is a 4.3-ms pulsar in a 22.2-day orbit with a companion that is likely to be a low-mass WD.
The unusual characteristic of this system is its orbital eccentricity, $e \, = \, 0.0798$, which is
much larger than those of most MSP - WD systems. This system is not unique in this respect; there are
four other similar systems with orbital periods between 22 and 32 days and 
eccentricities of the order of
0.1 (PSR~J2234+0611, \citealt{2013ApJ...775...51D,2019ApJ...870...74S}, PSR~J1946+3417, \citealt{2013MNRAS.435.2234B,2017MNRAS.465.1711B},
PSR~J0955$-$6150, \citealt{2015ApJ...810...85C} and PSR~J1618$-$3921, \citealt{2018A&A...612A..78O}).
Such similarities are not expected from a chaotic process
like the triple disruption that is
thought to have formed the unusual eccentric MSP - main sequence star PSR~J1903+0327
\citep{2008Sci...320.1309C,2011MNRAS.412.2763F}, however, as we show later, the exact
formation mechanism for these binaries is still unknown. 

For PSR~J1950+2414, \cite{2015ApJ...806..140K} measured the rate of advance of
periastron, $\dot{\omega} \, = \, 0.0020(3)^\circ \rm yr^{-1}$. Assuming this is solely an effect
of general relativity, they estimated that the implied total mass of the system, $M$, is $2.3(4)\, M_{\odot}$.
No other PK parameters were measured, so it was not possible to separate the component
masses. Although the measurement of $\dot{\omega}$ is highly significant, the uncertainty on the
resulting $M$ was too large to draw any interesting conclusions about the system.

\subsection{Motivation and structure of the work}

In this work, we present the results of continued timing of these two binary systems. The main
aim of this project was to measure the proper motions of the two systems more precisely and
to improve (as in the case of PSR J1949+3106) or to obtain (as in the case of
PSR~J1950+2414) masses for the MSPs and their companions. 

Measuring NS masses is important for several reasons. First, their measurement allows, in some
cases, precise tests of nature of gravitational waves
\citep{2012MNRAS.423.3328F} and of the strong equivalence principle \citep{2018Natur.559...73A},
which represent stringent tests of general relativity (GR) and alternative theories of gravity. Large NS masses,
as in the case of PSR~J0348+0432 \citep{2013Sci...340..448A} and PSR~J0740+6620 \citep{2019arXiv190406759T}
introduce stringent constraints on the equation of state for super-dense matter, a fundamental
problem in nuclear physics and astrophysics (see e.g., \citealt{2016ARA&A..54..401O} and references
therein). Apart from this, measuring more NS masses is important for understanding the
relation between the masses of the NS components and the orbital and kinematic properties of the
systems, which are a product of supernova physics (see e.g., \citealt{2017ApJ...846..170T} and references therein). 
Measuring the masses of NSs in PSR - NS and PSR - Massive WD systems,
where there was little accretion, is important for establishing the distribution of
NS birth masses (see e.g., \citealt{2017ApJ...844..128C}
and references therein). Finally, measuring MSP masses is important for understanding the role that
strong recycling (with potentially significant amounts of matter being accreted) have on
the observed mass distribution. In this respect, it is important to determine whether the
observed MSP mass distribution is uni- or bi-modal \citep{2016arXiv160501665A}. 

Many of these applications require an improvement in the statistics of well-measured NS masses, and some
require specifically an increase in the number of precise MSP mass measurements.
Our initial analysis of the two pulsars studied in this paper found that,
with adequate timing data, they would yield good mass measurements; this
initial expectation was, as we show below, largely confirmed.

\begin{figure}
\begin{center}
\includegraphics[width=\columnwidth, angle=0]{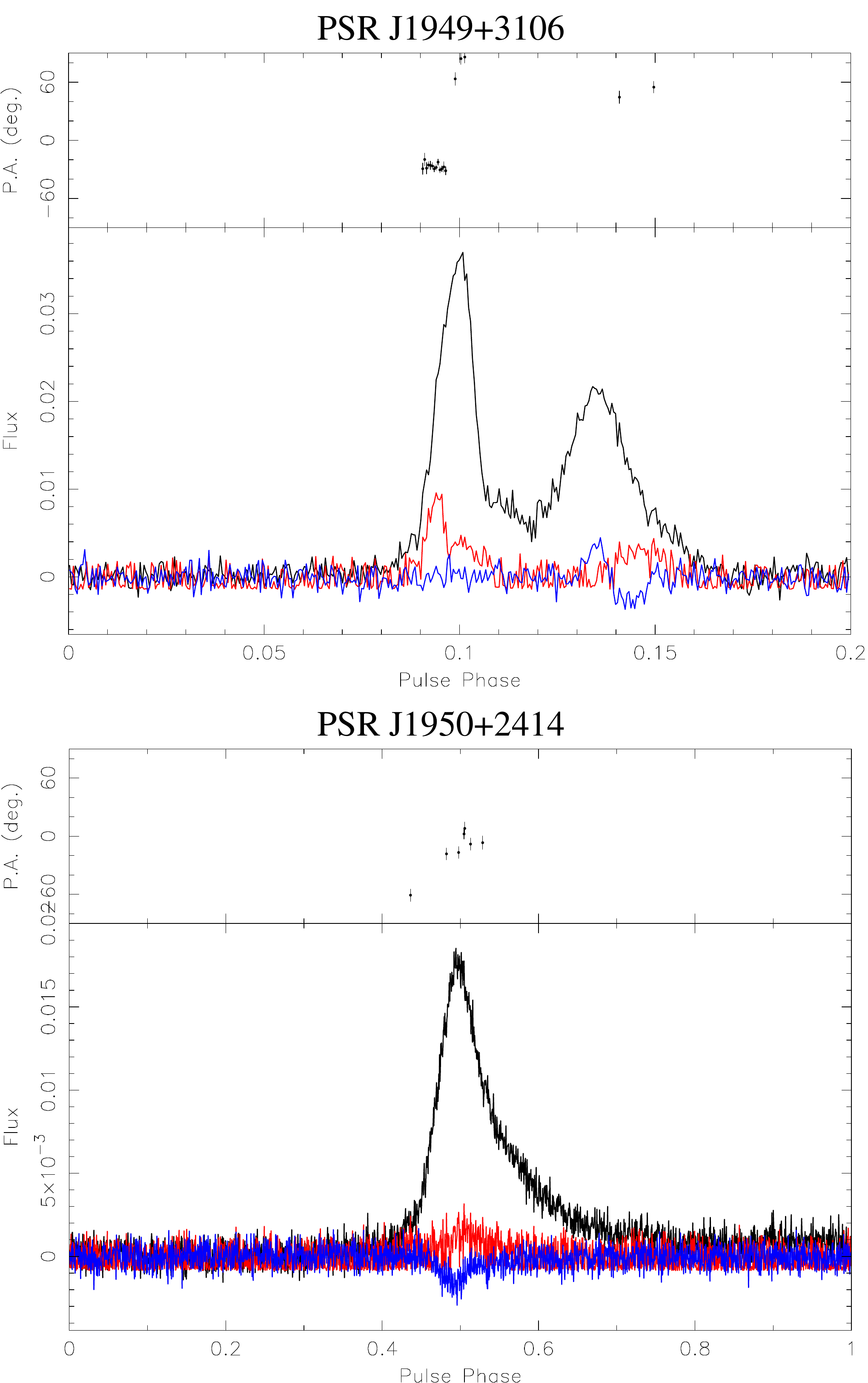}
\end{center}
\caption{Polarization profiles for PSR~J1949+3106 (top, zoomed in on
spin phase between 0 and 0.2) and PSR~J1950+2414 (bottom) obtained
with the Arecibo L-band receiver (1150 - 1730 MHz).
For the latter pulsar, the effects of scattering are evident.
The black lines indicate the total intensity, the red lines
the amount of linear polarization, and in blue the amount of
circular polarization.
The panels above each profile show the position angle (PA) of
the linear polarization.
}
\label{figure:profiles}
\end{figure}

In section~\ref{sec:observations}, we will present the new observations
we have made for this project and describe briefly how was the resulting data
reduced. In section~\ref{sec:results}, we present the timing results, with
a particular focus on the proper motion and the mass measurements. Finally, in
section~\ref{sec:discussion}, we discuss the
implications of these measurements for the origin and evolution of these systems. 

\begin{table*}
\begin{center}{\scriptsize
\caption{Ephemerides for two binary pulsars\label{tab:solutions}}
\begin{tabular}{l c c c c}
\hline
PSR & \multicolumn{2}{c}{J1949+3106} & \multicolumn{2}{c}{J1950+2414} \\
  \hline
  \multicolumn{5}{c}{Observation and data reduction parameters}\\
  \hline
  Reference epoch (MJD) \dotfill                               & \multicolumn{2}{c}{56000} & \multicolumn{2}{c}{56000} \\
  Span of timing data (MJD) \dotfill                           & \multicolumn{2}{c}{54389-57887} &
\multicolumn{2}{c}{55838-57883}  \\
  Number of ToAs \dotfill                                      & \multicolumn{2}{c}{3441} &
\multicolumn{2}{c}{913}  \\
  Assumed Solar wind parameter, $n_0$ (cm$^{-3}$) \dotfill                             & \multicolumn{2}{c}{0} & \multicolumn{2}{c}{6}  \\
  RMS Residual ($\mu s$) \dotfill                              & 3.8 & 3.8 & 4.7 & 4.7  \\
  $\chi^2$ \dotfill                              & 3629.6 & 3629.7 & 908.4 & 908.6  \\
  Reduced $\chi^2$ \dotfill                              & 1.063 & 1.063 & 1.023 & 1.023  \\
  \hline
  \multicolumn{5}{c}{Astrometric and spin parameters}\\
  \hline
  Right Ascension, $\alpha$ (J2000) \dotfill                   & 19:49:29.6374604(37) & 19:49:29.6374604(37)  & 19:50:45.063662(24) & 19:50:45.063661(24)\\
  Declination, $\delta$ (J2000) \dotfill                       & 31:06:03.80604(7) & +31:06:03.80604(6) & +24:14:56.9639(4) & +24:14:56.9639(4) \\
  Proper motion in $\alpha$, $\mu_{\alpha}$ (mas yr$^{-1}$) \dotfill & $-$2.895(31) & $-$2.894(31) & $-$2.12(18) & $-$2.12(18) \\
  Proper motion in $\delta$, $\mu_{\delta}$ (mas yr$^{-1}$) \dotfill & $-$5.091(46) & $-$5.093(45) & $-$3.63(19) & $-$3.64(19) \\
  Spin frequency, $\nu$ (Hz) \dotfill						   & 76.1140249316118(7) & 76.1140249316119(6) & 232.300152212241(37) & 232.300152212240(37) \\
  Spin frequency derivative, $\dot{\nu}$ ($10^{-15}\, \rm Hz \, s^{-1}$) \dotfill & $-$0.544131(7) & $-$0.544131(7) & $-$1.01478(37) & $-$1.01477(37) \\
  Dispersion measure, DM ($\rm cm^{-3}\, pc$) \dotfill              & 164.1266824) & 164.12669(24) & 142.0842(31) & 142.0842(31)  \\
  Rotation measure, RM ($\rm cm^{-3}\, pc$) \dotfill              & 230(20) &  230(20) & - & - \\
  \hline
  \multicolumn{5}{c}{Binary parameters}\\
  \hline
  Orbital model \dotfill                                        & DDFWHE & DDGR & DDFWHE & DDGR \\
  Orbital period, $P_{\rm b}$ (days) \dotfill                   & 1.9495374(6) & 1.94953755(20) & 22.19137127(6) & 22.19137127(6) \\
  Projected semi-major axis, $x$ (lt-s) \dotfill 				& 7.2886512(13) & 7.2886510(7) & 14.2199356(8) & 14.21993591(35) \\
  Epoch of periastron, $T_0$ (MJD) \dotfill                     & 56000.58744(47)  & 56000.58740(44) & 56001.3656142(26) & 56001.3656143(26) \\
  Orbital eccentricity, $e$ \dotfill                            & 0.000043124(36) & 0.000043122(35) & 0.07981175(6) & 0.07981173(4) \\
  Longitude of periastron, $\omega$ ($^\circ$) \dotfill         & 207.87(9) & 207.86(8) & 274.47042(4) &  274.47042(5) \\
  Total mass, $M$ ($\Msun$ ) \dotfill                           & 1.99$^{a}$ & 2.124(20) & $1.7790^{a}$ & 1.779(25) \\
  Companion mass, $M_c$ ($\Msun$ ) \dotfill                     & 0.81$^{b}$ & 0.81(5) & $0.300^{b}$  & 0.2788(38) \\
  Rate of advance of periastron, $\dot{\omega}$ ($^\circ \rm yr^{-1}$) \dotfill & 0.103(19) & - & 0.001678(16) & - \\
  Orthometric amplitude of Shapiro delay, $h_3$ ($\mu$s) \dotfill & 2.33(7) & - & 0.71(12) & - \\
  Orthometric ratio of Shapiro delay, $\varsigma$ \dotfill      & 0.837(12) & - & $0.783346^{c}$ & - \\
  Derivative of $P_{\rm b}$, $\dot{P}_{\rm b}$ ($10^{12}$ s s$^{-1}$) \dotfill 		& $-$0.046(28) & $-0.037(25)^{d}$ & $-1(11)$ & $0(11)^{d}$ \\
  \hline
  \multicolumn{5}{c}{Derived parameters}\\
  \hline
  Galactic longitude, $l$ \dotfill                              & 66.8583 & 66.8583 & 61.0975 & 61.0975 \\
  Galactic latitude, $b$ \dotfill                               & 2.5536 & 2.5536 & $-$1.1687 & $-$1.1687 \\
  DM-derived distance, $d_{\rm psr, 1}$ (kpc) \dotfill                       & 6.5 & 6.5 & 5.6 & 5.6  \\
  DM-derived distance, $d_{\rm psr, 2}$ (kpc) \dotfill                       & 7.5 & 7.5 & 7.3 & 7.3  \\
  Galactic height, $Z_{\rm psr, 1}$ (kpc) \dotfill                           &  0.29  & 0.29  &  0.11  & 0.11 \\
  Galactic height, $Z_{\rm psr, 2}$ (kpc) \dotfill                           &  0.33  & 0.33  &  0.15  & 0.15 \\
  Magnitude of proper motion, $\mu$ ($\rm mas \, yr^{-1}$) \dotfill                 &  5.856(43)  & 5.858(42) &  4.21(19)  & 4.21(19) \\
  Heliocentric transverse velocity, $v_{\rm T}$ ($\rm km \, s^{-1}$) \dotfill                 &  208(31)  & 208(31)  &  145(22)  & 145(22) \\
  Position angle of proper motion, $\Theta_{\mu}$ ($\deg$, J2000) \dotfill                 & 209.62(34)  & 209.60(34)  &  210.2(2.5)  & 210.2(2.5) \\
  $\Theta_{\mu}$ ($\deg$, Galactic) \dotfill                 & 268.69(34)  & 268.67(34)  &  269.2(2.5) & 269.1(2.5) \\  
  Pulsar spin period, $P$ (ms) \dotfill                            &  13.13818315216540(12) & 13.13818315216539(11) & 4.3047754832565(7) & 4.3047754832565(7) \\
  Spin period derivative, $\dot{P}$ ($10^{-20}$~s s$^{-1}$) \dotfill  & 9.39235(13) & 9.39235(13) & 1.8805(7) & 1.8805(7) \\
  Intrinsic spin period derivative, $\dot{P}_{\rm int}$ ($10^{-20}$~s s$^{-1}$) \dotfill  &  $9.40_{-0.78}^{+0.48}$ & $9.40_{-0.78}^{+0.48}$ & $2.033_{-0.014}^{+0.003}$ & $2.033_{-0.014}^{+0.003}$ \\
  Surface magnetic flux density, $B_0$ ($10^{9}$ Gauss) \dotfill & 1.12 & 1.12 & 0.30 & 0.30 \\
  Characteristic age, $\tau_c$ (Gyr) \dotfill                   &  2.2 & 2.2 & 3.4 & 3.4  \\
Spin-down power, $\dot{E}$ ($10^{33}\, \rm erg \, s^{-1}$) \dotfill                   &  1.6 & 1.6 & 10.1 & 10.1  \\
  Mass function, $f$ ($\Msun$ ) \dotfill                        & 0.10938583(8) & 0.10938582(5) & 0.0062691312(11) & 0.0062691316(5) \\
  Pulsar mass, $M_{p}$ ($\Msun$ ) \dotfill                      & - & 1.33(15) & - & 1.500(22)  \\
  Orbital inclination ($\deg$) \dotfill                 & 79.9 & 79.9 & 76.1 & 76.1 \\
  \hline
\multicolumn{5}{l}{Notes. Timing parameters derived using {\sc tempo}, they are derived in the Barycentric Dinamical Time (TDB), using DE 436 Solar System ephemeris.}\\
\multicolumn{5}{l}{{\it a}: Derived from $\dot{\omega}$, {\it b}: derived from $h_3$ and $\varsigma$, {\it c}: assumed from best inclination given by DDGR fit, {\it d}: Fitted as the XPBDOT parameter.}\\
\multicolumn{5}{l}{$d_1$ is derived using the NE2001 \citep{2002astro.ph..7156C} Galactic model, $d_2$ using the YMW16 \citep{2017ApJ...835...29Y} Galactic model.}\\
\multicolumn{5}{l}{Estimate of $v_{\rm T}$, $\dot{P}_{\rm int}$ assumes $d_2$ with an uncertainty of 15 \%.}
\end{tabular}}
\end{center}
\end{table*}

\begin{figure*}
\begin{center}
\includegraphics[width=\textwidth, angle=0]{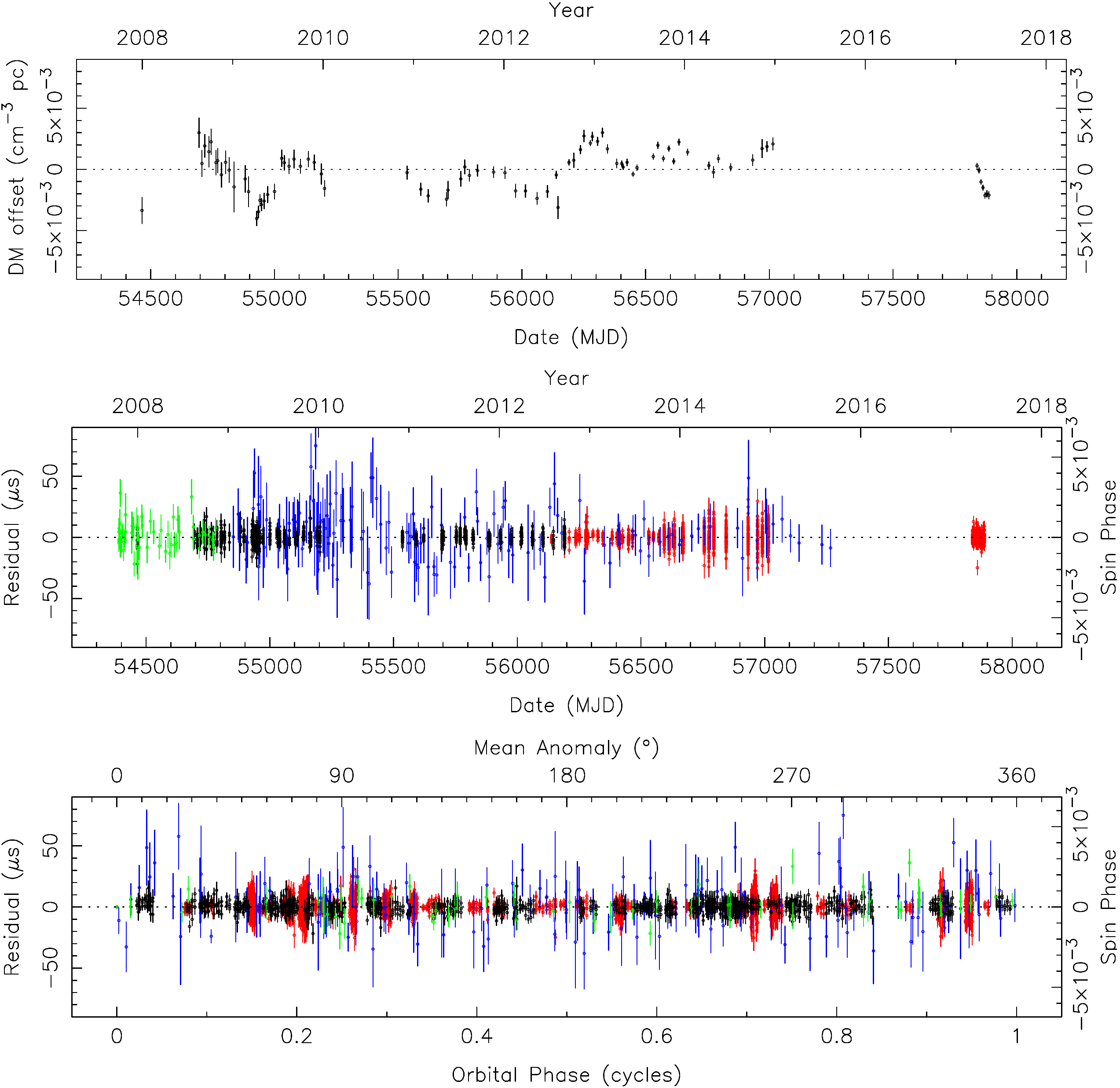}
\end{center}
\caption{{\em Top:} Dispersion measure offsets of PSR J1949+3106 relative
to the DM fixed in Table~\ref{tab:solutions} as a function of date; these
are obtained using a DMX-type ephemeris.
We only present the DM offsets of time segments with
TOAs measured at multiple frequencies; these were only derived for Arecibo data.
{\em Middle:} post-fit ToA residuals for all TOA data sets using 
the DDGR ephemeris in Table~\ref{tab:solutions}
(which used DM derivatives, {\em not} the DMX parameterization) as a function of date and
{\em Bottom:} as a function of
orbital phase. The residual 1-$\sigma$ uncertainties are indicated by
vertical error bars. Black indicates data taken with the WAPP correlators,
red indicates data taken with PUPPI, blue indicates data taken at
Jodrell Bank with the ROACH system and green data taken with the GBT.
}
\label{figure:residuals_J1949+3106}
\end{figure*}

\begin{figure*}
\begin{center}
\includegraphics[width=\textwidth, angle=0]{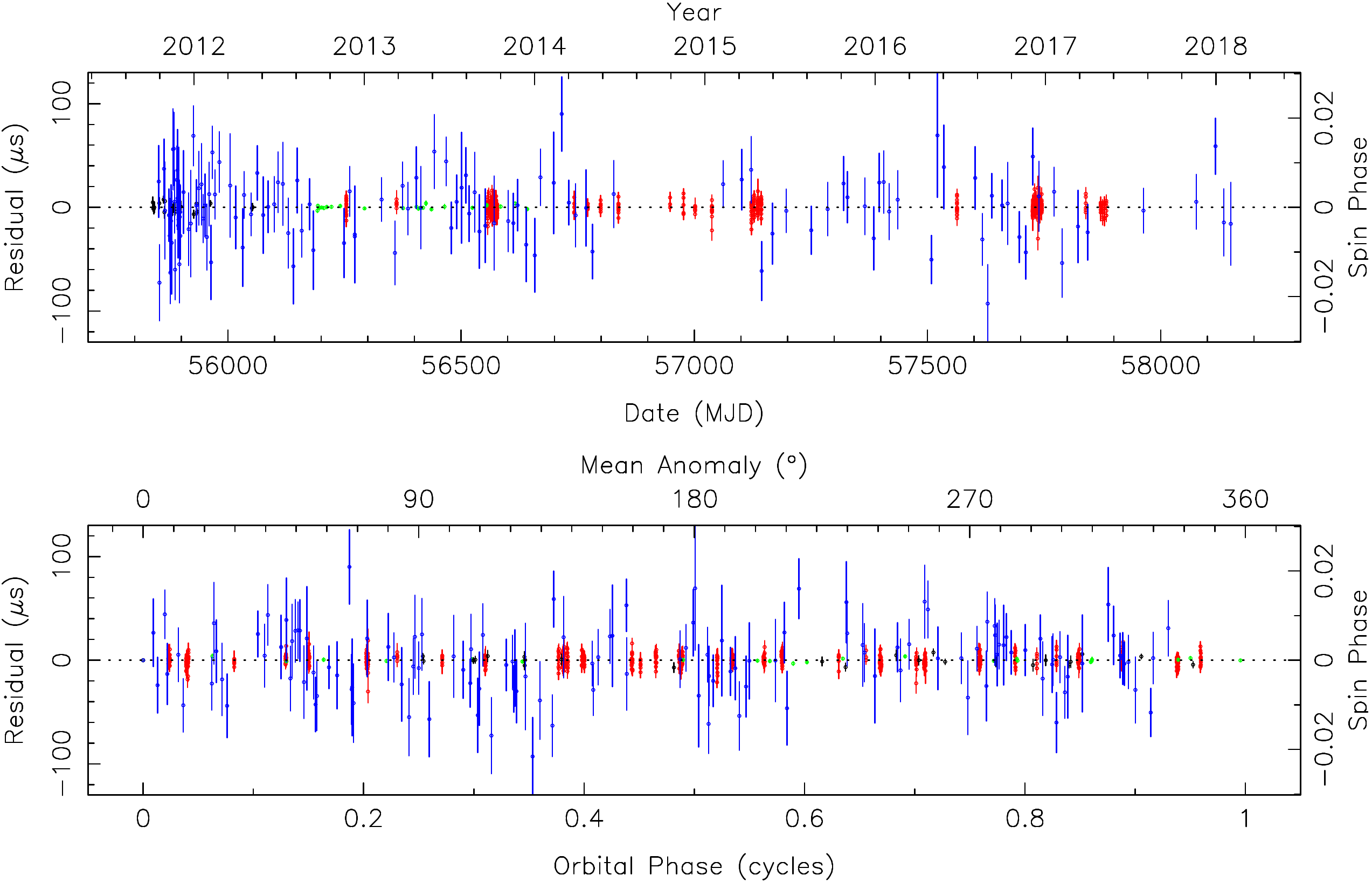}
\end{center}
\caption{{\em Top:} ToA residuals for the DDGR ephemeris of PSR J1950+2414 in
Table~\ref{tab:solutions} as a function of date and
{\em Bottom:}
orbital phase. The residual 1-$\sigma$ uncertainties are indicated by
vertical error bars. Black indicates data taken with the Mock spectrometers,
green indicates data taken with PUPPI in incoherent mode, red data taken with
PUPPI in coherent mode and blue indicates data taken at
Jodrell Bank with the ROACH system.
}
\label{figure:residuals_J1950+2414}
\end{figure*}

\section{Observations and data reduction}
\label{sec:observations}

\subsection{Observations}

For both systems, we re-use the topocentric pulse times of arrival (ToAs)
used in their published timing; for detailed descriptions
see \citep{2012ApJ...757...89D} for PSR J1949+3106 and \citep{2015ApJ...806..140K} for PSR~J1950+2414.
For PSR~J1949+3106, the data set is dominated by Arecibo data taken with the
Wideband Arecibo Pulsar Processors (WAPPs, \citealt{2000ASPC..202..275D}). For PSR~J1950+2414,
the setup was similar to that used during the PALFA survey (which uses the 
Mock spectrometers as a back-end), but on dedicated timing campaigns the
Puerto Rico Ultimate Pulsar Processing Instrument (PUPPI) was used, mostly with the 
``L-wide'' receiver, which is sensitive to radio frequencies between 1170 to 1730 MHz.

Most of the new data on both systems was taken with PUPPI in coherent dedisperison mode
and the L-wide, in the way described by \cite{2015ApJ...806..140K}.
All PUPPI data for both pulsars was processed as independent 100-MHz blocks, the ToAs
are derived from the integrated pulse profiles within each 100-MHz block using the standard
{\tt PSRCHIVE}\footnote{\url{http://psrchive.sourceforge.net/index.shtml}} \citep{2004PASA...21..302H}
routines as described in the aforementioned works. For PSR J1949+3106, we used separate
profile templates for each 100-MHz sub-band to derive ToAs that account for any frequency-dependent evolution of the profile shape \citep[This method was earlier used by][where further details can be found.]{2019A&A...624A..22D} This has significantly improved the timing precision
of this pulsar relative to the standard single-template method; part of the
reason is the strong profile evolution for this pulsar as a function of frequency \citep{2012ApJ...757...89D}.
For PSR J1950+2414, this method did not improve the timing noticeably, so we used
TOAs derived from the standard single-template method.

For PSR~J1949+3106, more than half of the additional PUPPI data resulted from the
North American Nanohertz Observatory for Gravitational Waves (NANOGrav)
observations \citep{2018ApJS..235...37A}, these were taken between
MJDs 56139 and 57015 (2012 July 31 to 2014 December 23),
which were used to test the suitability of this pulsar for pulsar timing arrays (PTAs);
the pulsar was eventually dropped out of the PTA. To this, we added the ToAs from
a dense, targeted orbital campaign that happened between MJDs 57839 and 57887 (2017 March 27 and May 14).

For PSR J1950+2414, most of the additional data were obtained during two dense orbital campaigns, 
the first between MJDs 56557 and 56576 (2013 September 22 to October 11) and the second between
MJDs 57725 and 57744 (2016 December 3 to 22), with more sparse observations made at
other times.

The ToA analysis is made using {\tt tempo}\footnote{\url{http://tempo.sourceforge.net/} \citep{2015ascl.soft09002N}}.
To convert the telescope ToAs (corrected to the International Bureau of Weights and Measures
version of Terrestrial Time, TT) to the Solar System barycentre, we used the
Jet Propulsion Laboratory's DE436 solar system ephemeris; the
resulting timing parameters are presented in Barycentric Dynamical Time (TDB).

The timing solutions for both pulsars are presented in Table~\ref{tab:solutions}.
All uncertainties are as derived by {\sc tempo}, and quoted to 1$\sigma$ (68.3\% C. L.).
For each pulsar, we present two solutions, one based on the DDGR orbital model \citep{1986AIHS...44..263D},
which assumes the validity of GR to derive self-consistent mass values,
and a second solution based on a theory-independent parameterization of the
relativistic effects observed in the timing of these systems, the DD model
\citep{1986AIHS...44..263D} with a re-parameterization of the Shapiro delay known as the
``orthometric'' parameterization \citep{2010MNRAS.409..199F}. Its implementation
in {\tt tempo} has the name DDFWHE \citep{2016ApJ...829...55W}.
The ToA residuals obtained with the DDGR solutions for PSRs~J1949+3106 and J1950+2414
are presented in Figures~\ref{figure:residuals_J1949+3106}
and \ref{figure:residuals_J1950+2414}, respectively.

We modeled the DM by introducing DM derivatives; 7 in the case of PSR~J1949+3106
and 6 in the case of J1950+2414.
The number of derivatives was determined by the significance of the improvement,
we stopped adding new derivatives when the $\chi^2$ improvement
becomes smaller than 2.

For PSR J1949+3106, we investigated the possibility of using the
DMX model, which fits for a time-varying DM \citep{2013ApJ...762...94D}.
In this model we selected ranges of 6 days for independent DM estimates, as
in the NANOGrav data analysis \citep{2018ApJS..235...37A};
the measured DM values are presented in the top plot of
Figure~\ref{figure:residuals_J1949+3106}, as a function of time.
We only present the DM offsets of time segments with
TOAs measured at multiple frequencies; these were only derived for Arecibo data,
where we obtained independent TOAs for every 100 MHz sub-band of the L-wide data.
The problem of this approach is that the mass values are strongly
dependent of the DMX interval we choose to use. This stems,
we believe, from the fact that we only have data taken at L-band.
Multi-receiver data would be necessary to make more robust measurements
of the DM variations in a way that the uncertainties do not affect the
small time signatures from Shapiro delay.

\section{Results}
\label{sec:results}

\subsection{Proper motions}

For both systems we can measure precise proper motions, with total magnitude
$\mu$ and a position angle $\Theta_\mu$ given in Table~\ref{tab:solutions},
both in Equatorial (J2000) and Galactic coordinates. For both pulsars,
this has some covariance with the DM variations, for
this reason the proper motion and velocity uncertainties are under-estimated.

Both systems have Westwards proper motions along the Galactic plane.
The vertical velocities (i.e., the Heliocentric velocities in the
direction perpendicular to the Galactic plane) are small: 
$v_{V} = \, \mu \cos(\Theta_{\mu}) d_{\rm psr}\, = \, (-0.63 \pm \, 0.16) \, \times\, d_{\rm J1949+3106}\, \rm km \, s^{-1}$ and
$v_{V} \, = \, (-0.3 \pm \, 0.9) \, \times\, d_{\rm J1950+2414}\, \rm km \, s^{-1}$
where $d_{\rm psr}$ are their distances.
We cannot measure the parallaxes from the timing
with any significance, so we derive the distances
from their DMs using the NE2001 \citep{2002astro.ph..7156C} and
the YMW16 \citep{2017ApJ...835...29Y} models of the Galactic electron distribution;
these are presented in table~\ref{tab:solutions}. 
Assuming the YMW16 distances for both systems, their vertical velocities are
$-4.8\, \pm \, 1.2$ and $-2.3 \pm 6.6 \, \rm km \, s^{-1}$
respectively.  The vertical velocity of the Sun in the Galaxy is
about $+7.3(1.0) \, \rm km \, s^{-1}$ \citep{2010MNRAS.403.1829S}; this means
that all objects with no vertical velocity are seen, in the Helocentric frame,
with an opposite vertical velocity. Subtracting that from the observed
proper motions, we obtain, in the reference frame of the Galaxy,
$v_V \, = \, 2.5\, \pm\, 1.2$ and $5.0\, \pm 6.6 \, \rm km \, s^{-1}$ respectively.

The vertical velocities are a component of the pulsar's velocity relative
to the local standard of rest (LSR), known as {\em peculiar velocity}.
The fact that they are small suggests that the pulsars might be in the LSR.
We now test this hypothesis by calculating the magnitude of the proper motions
parallel the Galactic plane (the {\em horizontal} proper motions) these
pulsars should have if they were in the LSR.

The first step is to calculate the X, Y and Z coordinates of the pulsar,
which is done easily enough from $l$, $b$, $d$ and the Sun's distance to
the Galactic centre, $r_0$. For this, we used the estimate 
from the GRAVITY experiment \citep{2018A&A...615L..15G}, $r_0 \, = \, 8.122(31) $kpc.
The coordinates for the Sun are $X_{\odot}\, =\, r_0$, $Y_{\odot}\, =\,  0$ and $Z_{\odot}\, \simeq \, 0.02$; for the
pulsar $X_{\rm psr}\, = \, r_0 - d_{\rm psr} \cos(l) \cos (b)$, $Y_{\rm psr}\, = \, d_{\rm psr} \sin(l) \cos (b)$
and $Z_{\rm psr}\, = Z_{\odot} + \, d_{\rm psr} \sin(b)$.

The Sun's peculiar velocity is given by $V_{X, \odot} \,=\, -11.1(1.5)$, $V_{Y, \odot} \,=\, 12.2(2.0)$
and $V_{Z, \odot} \,=\, 7.3(1.0)\, \rm km \, s^{-1}$ (here the X-direction is {\em away} from the
centre of the Galaxy, the opposite of the convention used by \citealt{2010MNRAS.403.1829S},
which define the X axis as pointing to the centre of the Galaxy). To get the velocity of the Sun relative
to the Galactic centre, we add the Galaxy's rotational velocity to $V_{Y, \odot}$.
Here we use the value provided by \cite{2018arXiv180809435M},
$v_{\rm Gal} \, = \, 233.3 \, \rm km \, s^{-1}$, which
already takes the updated $r_0$ into account.

If the pulsar is in the LSR, then $V_{X, \rm psr} = - v_{\rm Gal} Y_{\rm psr}/r_{\rm psr}$ and
$V_{Y, \rm psr} = v_{\rm Gal} X_{\rm psr}/r_{\rm psr}$, where $r_{\rm psr}$ is the pulsar's
distance from the Galactic centre. Finally, we calculate the projection of the
difference of velocities along a unit vector perpendicular to the line of sight
and parallel to the plane of the Galaxy, and then divide the resulting velocity
by the distance to the pulsar to obtain the horizontal proper motion $\mu_H$.

For PSR~J1949+3106 ($l \, = \, 66.8583 \, \deg$) the NE2001 distance is 6.5 kpc,
and $r_{\rm psr} \, = \, 8.18\,$kpc, a distance very similar to $r_0$. 
From this we obtain $\mu_H \, = \, -5.85 \, \rm mas\, yr^{-1}$, where
the negative sign indicates Westwards motion along the Galactic plane (thus,
decreasing $l$).
For the YMW16 distance (7.5 kpc), $r_{\rm psr} \, = \, 8.60\,$kpc and $\mu_H \, = \, -5.69 \, \rm mas\, yr^{-1}$.
The observed $\mu_H$, given by $\mu \sin(\Theta_{\mu}) = -5.86(4) \, \rm mas\, yr^{-1}$, is
in very good agreement with the NE2001 estimate, and in 4.3-$\sigma$
disagreement with the YMW16 estimate.
If the NE2001 distance is correct, the peculiar horizontal velocity of the
pulsar is $v_H \, = \, 0.0(1.3) \, \rm km \, s^{-1}$;
if the YMW16 model distance is correct
$v_H \, = \, 5.9(1.5) \, \rm km \, s^{-1}$.  Thus the precisions of $v_H$ is limited
by the uncertainty of the distance. If we assume that the pulsar is in the LSR, then the
NE2001 distance estimate is much closer to the real distance than the YMW16 
estimate.

For PSR~J1950+2414 ($l \, = \, 61.0975 \, \deg$) the NE2001 distance is 5.6 kpc,
thus $\mu_H \, = \, -6.11 \, \rm mas\, yr^{-1}$.
For the YMW16 distance (7.3 kpc) $\mu_H \, = \, -6.04 \, \rm mas\, yr^{-1}$.
The observed horizontal proper motion, $\mu_H \, = \, -4.21(19) \, \rm mas\, yr^{-1}$, is
in clear disagreement with the estimates above, the differences
are $+1.90(19)$ and $+1.83(19) \, \rm mas\, yr^{-1}$;
this is about 10-$\sigma$ significant in both cases. Using the YMW16
distance, we obtain $v_H \, = \, +61(7) \, \rm km \, s^{-1}$. Thus PSR~J1950+2414
has a significant horizontal peculiar velocity, despite its very small vertical velocity.

\subsection{DM evolution}

If PSR~J1949+3106 were in the LSR, and if $r_{\rm psr} \, \simeq \, r_0$
(this happens if the pulsar is near the NE2001 distance), 
then something interesting would happen: The pulsar and the Sun would not only travel
with nearly the same absolute velocity as all objects in the LSR ($v_{\rm gal}$),
but they would also have approximately the same
angular velocity ($\Omega_{\rm gal} = v_{\rm gal} / r_0$) around the
Galactic centre. In such a configuration, the
Sun and PSR~J1949+3106 would form an approximately rigid
rotating triangle, i.e., one with approximately constant side length.
Thus the distance between the
Sun and PSR~J1949+3106 would change little, i.e., J1949+3106
should have a very small Heliocentric radial velocity. A detailed calculation
yields a radial velocity of $-16.9 \, \rm km \, s^{-1}$ that is
mostly caused by the Sun's peculiar velocity, without the latter the radial
velocity would be $-1.3 \, \rm km \, s^{-1}$.

Currently, it is not possible to measure the Heliocentric radial velocity of
this system: the quantity cannot be extracted from pulsar timing.
It could in principle be extracted from absorption lines
in the optical spectrum of the companion,
however, the companion to PSR~J1949+3106 has not been detected.
For this reason, we cannot measure the peculiar velocity of PSR~J1949+3106
along the radial direction, so we cannot conclude that the system
really has a low peculiar velocity; even though this is true in the
direction perpendicular to the line of sight.

Recently, \cite{2017ApJ...841..125J} used the observed DM variations
of the NANOGrav MSPs in an attempt to measure the radial velocities of the
pulsars. A detailed calculation of all effects is beyond the scope of
this work. However, if the pulsar and the intervening ionized interstellar medium (IISM) 
are all in the LSR, then there should be very little change in the
line of sight of the system relative to the IISM, since the latter is 
basically co-moving with the Earth and the system. This means
that there should be no long-term increasing or decreasing 
trend in the DM of the pulsar. This is in agreement with the observations
presented in Fig.~\ref{figure:residuals_J1949+3106} -- the DM derivative is $0.00060(20) \, \rm cm^{-3} \, pc \, yr^{-1}$, 3-$\sigma$ consistent with 
zero -- and in disagreement with the DM trends observed for most other
pulsars listed by \cite{2017ApJ...841..125J}.

\subsection{Spin parameters}
\label{sec:spin}

With the assumed distances and the measured proper motion we can estimate the magnitude of the 
kinematic effects that may bias the timing parameters using the simple expressions
provided by \cite{1970SvA....13..562S} and \cite{1991ApJ...366..501D}, which
estimate the rate of change of the Doppler shift factor $D$:
\begin{equation}
\label{eq:Doppler}
\frac{\dot{D}}{D} \, = \, - \frac{a_l + \mu^2 d}{c}    
\end{equation}
To estimate the Galactic acceleration $a_l$, we use the equations provided by \cite{2009MNRAS.400..805L}.
These are reasonably accurate for these pulsars since they are very close to the Galactic plane.
In those equations we used the $r_0$ and $v_{\rm Gal}$ from the previous section.
The results for $\dot{D}/D$ are $0.5^{+3.6}_{-5.9} \, \times \, 10^{-20} \rm s^{-1}$ and
$35.6^{+0.6}_{-3.3} \, \times \, 10^{-20} \rm s^{-1}$ for J1949+3106 and J1950+2414 respectively,
both values calculated assuming the YMW16 distances.
The expected value for PSR~J1949+3106 (consistent with zero) is consistent with the
``rigid rotating triangle'' configuration mentioned above where the distance between
the pulsar and the Earth is unchanging: indeed, $\dot{D}$ is proportional to
the second derivative of that distance.

With the values of $\dot{D}/D$ for both pulsars, we can estimate the 
kinematic correction to the spin period derivatives, $\dot{P}_{\rm kin} \, = \, - P \dot{D}/D$.
Subtracting this from the observed $\dot{P}$s we obtain intrinsic spin period derivatives
($\dot{P}_{\rm int}$); these are quite
similar to the observed values for both pulsars. From $P$ and $\dot{P}_{\rm int}$,
we derive updated values for  the characteristic age ($\tau_c$), surface 
magnetic flux density ($B_0$) and also the rate of loss of rotational energy ($\dot{E}$)
using the standard expressions presented by \cite{2004hpa..book.....L}. All these quantities
are presented in Table~\ref{tab:solutions}.

\subsection{Orbital period derivatives}

The observed orbital period derivative for these binaries with compact companions is given,
to first order, by:
\begin{equation}
\left( \frac{\dot{P}_{\rm b}}{P_{\rm b}} \right)_{\rm obs} \, = \, \left( \frac{\dot{P}_{\rm b}}{P_{\rm b}} \right)_{\rm GW} \, - \, \frac{\dot{D}}{D},
\end{equation}
where the first term is the orbital decay caused by the emission of gravitational waves
and the second term is the fractional rate of change of the Dopper shift calculated
in eq.~\ref{eq:Doppler}.

In the case of PSR~J1950+2414, it is clear that only the second term matters.
Indeed, in that case the DDGR model predicts
$\dot{P}_{\rm b, GR}\, = \, -4.47 \, \times \, 10^{-17} \rm s\, s^{-1}$, this is
four orders of magnitude smaller than the kinematic term
$\dot{P}_{\rm b, k}\, \equiv \, - P_{\rm b} \dot{D}/D \, = \, -0.68_{-0.01}^{+0.06} \, \times \, 10^{-12} \rm s\, s^{-1}$. If we fit for this quantity, we obtain
$\dot{P}_{\rm b, obs}\, = \, -1 \pm 11 \, \times \, 10^{-12} \rm s\, s^{-1}$.
Thhis means we're one order of magnitude away from the detection of 
$\dot{P}_{\rm b, k}$.

For PSR~J1949+3106, the situation is far more interesting. In that case,
$\dot{P}_{\rm b, GR}\, \simeq \, -6 \, \times \, 10^{-15} \rm s\, s^{-1}$, is
comparable to the uncertainty of the kinematic term
$\dot{P}_{\rm b, k}\, = \, -0.8^{+10.0}_{-6.1} \, \times \, 10^{-15} \rm s\, s^{-1}$,
obtained assuming the YMW16 distance with a 15\% uncertainty. Our DDFWHE solution yields
$\dot{P}_{\rm b, obs}\, = \, -46 \pm 56 \, \times \, 10^{-15} \rm s\, s^{-1}$ (95.4 \% C. L.).
This precision is not yet enough to constrain the distance to the system.

\begin{figure*}
\begin{center}
\includegraphics[width=0.8\textwidth, angle=0]{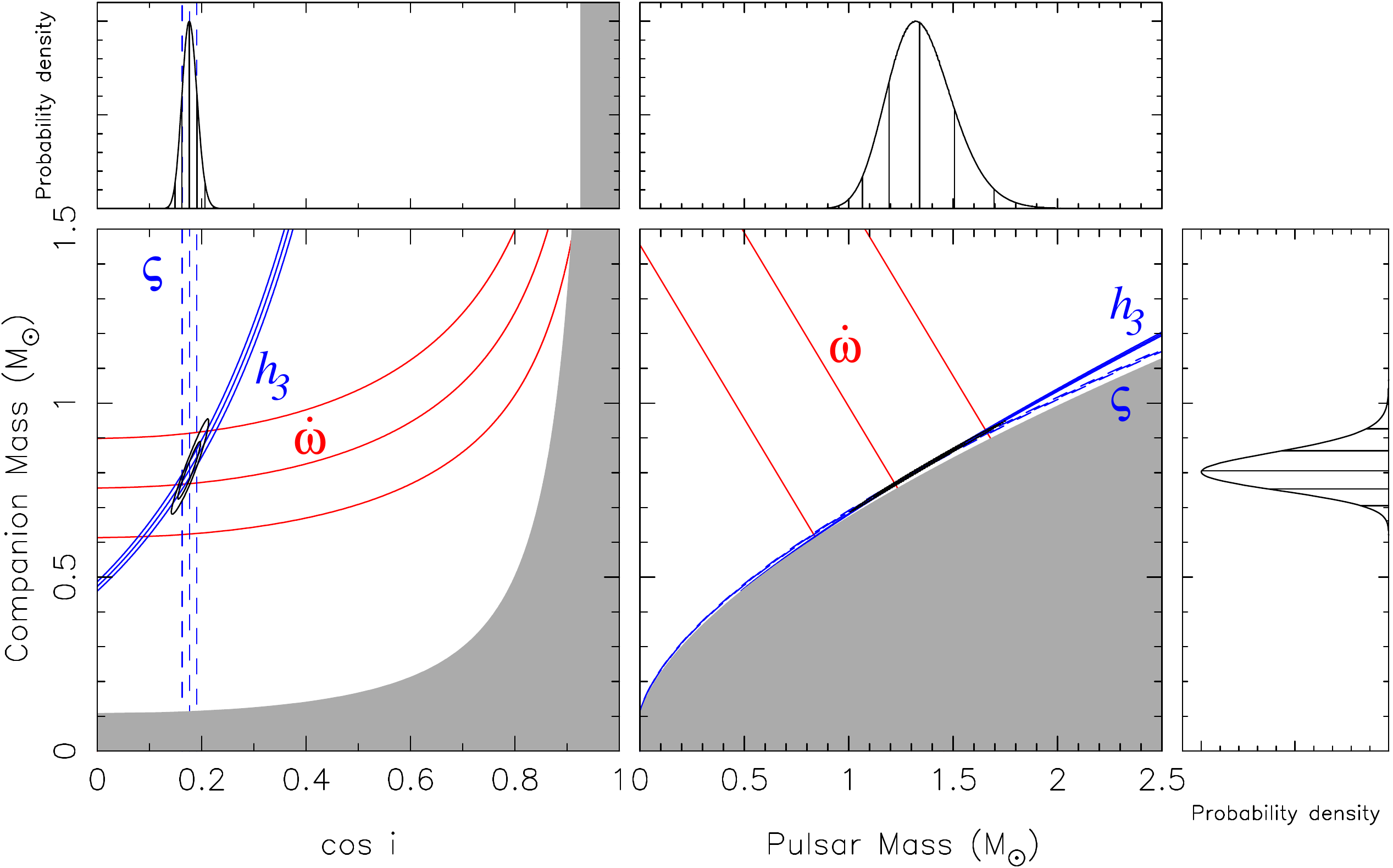}
\end{center}
\caption{Mass constraints for PSR~J1949+3106. In the main square plots, the
lines indicate the regions that are (according to general relativity)
consistent with the nominal and $\pm\, 1 \sigma$ measurements of $h_3$ (solid blue),
$\varsigma$ (dashed blue) and $\dot{\omega}$ (solid red) obtained from
the DDFWHE model (see Table~\ref{tab:solutions}).
The contour plots include 68.23 and 95.44\% of the total
2-dimensional probability density function (pdf), derived from the quality
($\chi^2$) of the {\sc tempo} fits using a DDGR model
to the ToA data set we have obtained for this pulsar.
The location of the regions of high probability is well described by the
$h_3$ and $\varsigma$ parameters and their uncertainties, with a small
influence from $\dot{\omega}$. 
In the left plot, we display
the cosine of the orbital inclination ($\cos i$, which has, for randomly
inclined orbits, a flat pdf) versus the companion mass ($M_c$); the gray
region is excluded because the pulsar mass ($M_p$) must be larger than 0.
In the right plot, we display $M_p$ versus $M_c$; the gray region is excluded
by the constraint $\sin i\, \leq\, 1$. The side panels display the 1-d pdfs
for $\cos i$ (top left), $M_p$ (top right) and $M_c$ (right). The vertical lines
in these pdfs indicate the median and the percentiles corresponding to
1 and 2 $\sigma$ around the median.}
\label{figure:mass_mass_J1949+3106}
\end{figure*}

\begin{figure*}
\begin{center}
\includegraphics[width=0.85\textwidth, angle=0]{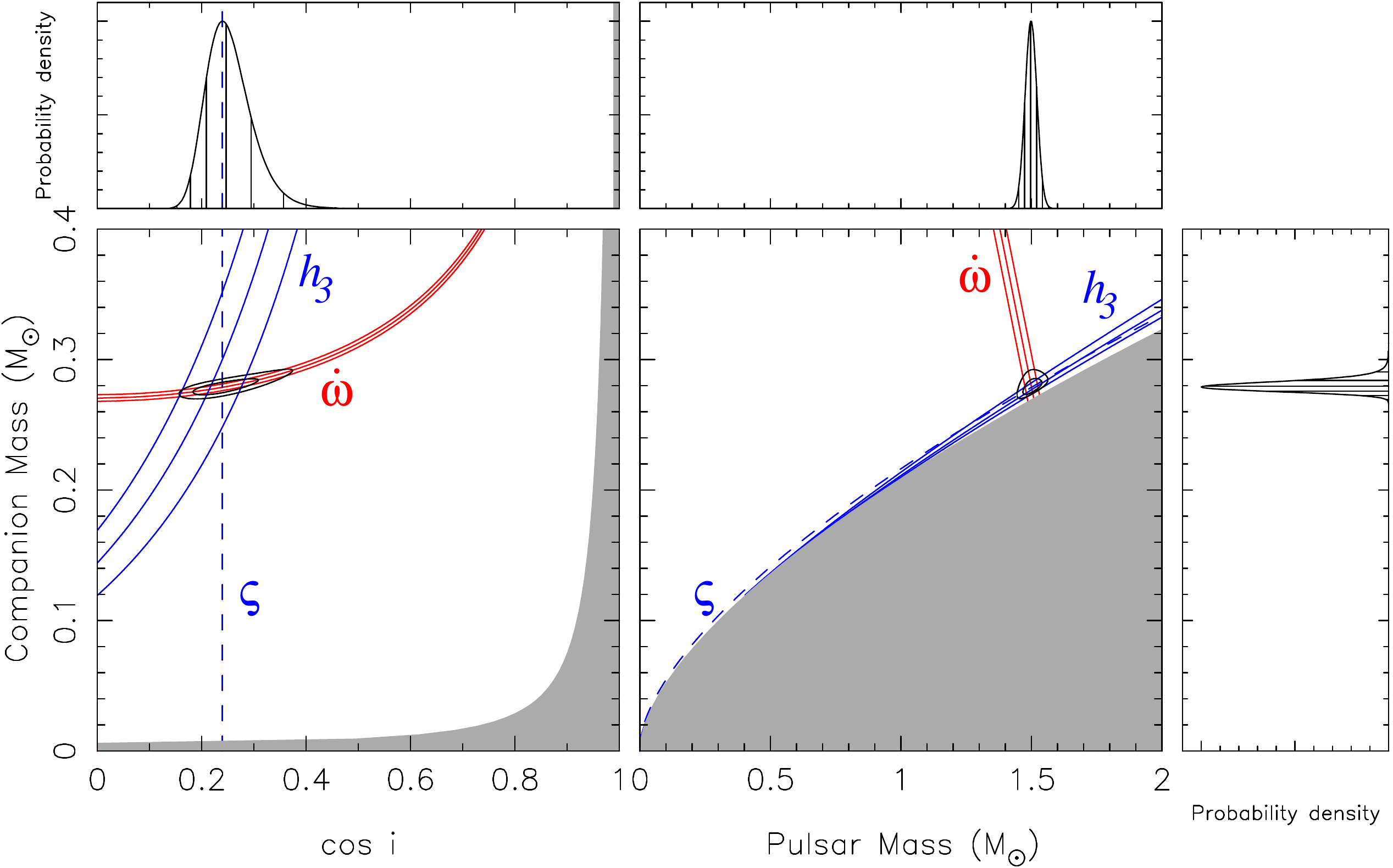}
\end{center}
\caption{Same as Fig.~\ref{figure:mass_mass_J1949+3106},
but this time for PSR~J1950+2414. One of the differences is the much larger uncertainties
for $h_3$ and $\varsigma$ described in the text; for that reason we fixed $\varsigma$
to the value predicted by the best DDGR ephemeris in Table~\ref{tab:solutions},
and then fitted for $h_3$ independently.
As in Fig.~\ref{figure:mass_mass_J1949+3106}, the probability density functions are
obtained with a fully self-consistent DDGR model. The regions of higher probability
are reasonably well described by the $\dot{\omega}$ and $h_3$ parameters and their
uncertainties.
}
\label{figure:mass_mass_J1950+2414}
\end{figure*}

\subsection{Mass measurements}

For each pulsar in this work, we detect 3 post-Keplerian parameters: the rate of
advance of periastron ($\dot{\omega}$), and the Shapiro delay, which yields
two parameters, the orthometric amplitude ($h_3$) and ratio ($\varsigma$).
The fits made using the DDGR model \citep{1986AIHS...44..263D} use the
relativistic time signatures of these effects and assume the
validity of GR to provide direct estimates of the total binary mass $M$
and the companion mass $M_c$. For PSR J1949+3106, these are respectively
$M \, = \, 2.14(20) \, M_{\odot}$, $M_c \, = \, 0.81(5) \, M_{\odot}$
and $M_p \, = \, 1.33(15) \, M_{\odot}$, given the mass 
function these correspond to $s\, =\, \sin i\, \simeq\, 0.9844...$.
These are 1-$\sigma$ consistent with the values
presented by \cite{2012ApJ...757...89D}, but $\sim \, 2$ times more precise.
These masses are based almost entirely on the measurement of the
Shapiro delay (see Fig.~\ref{figure:mass_mass_J1949+3106}), but are now being
influenced by the detection of $\dot{\omega}$.

One of the remarkable features of PSR~J1949+3106 is that
we have a 5-$\sigma$ significant measurement of $\dot{\omega}$
of this system despite the fact that its orbital eccentricity
is rather small ($e = \, \, 4.3124(36) \, \times \, 10^{-5}$). The
value we measure, $\dot{\omega} \, = \, 0.103(19) \deg \,\rm yr^{-1}$, is
consistent with the the GR prediction for the masses determined from the
Shapiro delay (see Fig.~\ref{figure:mass_mass_J1949+3106}). This is
a test of GR, although not a very constraining one.

For PSR~J1950+2414, the DDGR solution yields precise masses:
The total mass is $M \, = \, 1.779(25) \, M_{\odot}$, for
the individual masses we obtain: $M_c \, = \, 0.2788(38) \, M_{\odot}$
and $M_p \, = \, 1.500(22) \, M_{\odot}$.

The total mass is determined precisely because the unusually large
eccentricity of the system ($e \, = \, 0.07981175(6)$) enables a precise measurement
of the rate of advance of periastron ($\dot{\omega} \, = \, 0.001678(16)\, \deg \rm yr^{-1} $
in the DDFWHE solution). The constraints imposed by this, assuming 
that the effect is caused by GR alone, are shown 
with the red lines in Fig.~\ref{figure:mass_mass_J1950+2414}.

The Shapiro delay yields, on its own, far less precise masses
than for PSR J1949+3106; fitting for both $h_3$ and $\varsigma$
we get very low significance for both parameters.
To better estimate the regions allowed by the Shapiro delay,
we fix $\varsigma$ to the value of $\sin i$ derived by the DDGR solution
($s\, =\, \sin i\, = \, 0.9709...$), using
\citep{2010MNRAS.409..199F}:
\begin{equation}
\varsigma \, = \, \frac{s}{1 + \sqrt{1 - s^2}} = 0.7833...,
\end{equation}
this is represented by the blue dashed lines in figure~\ref{figure:mass_mass_J1950+2414}.
With this, we obtain a highly significant detection of the orthometric amplitude,
$h_3 \, = \, 0.71\, \pm\, 0.12\, \mu \rm s$. It is this constraint, in combination with the
$\dot{\omega}$, that allows the precise measurement of the individual component masses
(see figure~\ref{figure:mass_mass_J1950+2414}).

We now verify whether the $\dot{\omega}$ is caused by the effects of GR. To do this,
we estimate the contribution to $\dot{\omega}$
from the proper motion of the system ($\mu \, = \, 4.21(19) \, \rm mas \, yr^{-1}$),
using the equations first derived by \cite{1995ApJ...439L...5K,1996ApJ...467L..93K}.
The largest possible contribution is given by
\begin{equation}
    \dot{\omega}_{\rm k} \, = \, \pm \, \frac{\mu}{\sin i} \, = \, \pm \, 1.2 \, \times \, 10^{-6} \, \deg \rm yr^{-1},
\end{equation}
which is one order of magnitude smaller than the uncertainty of $\dot{\omega}$.
Other contributions are likely to be smaller \citep{2019ApJ...870...74S}.
Therefore, the assumption that the measured $\dot{\omega}$ is relativistic is,
to our knowledge, warranted.

\subsection{Bayesian mass estimates}

To verify the uncertainty estimates for the masses, we used the Bayesian method
described in detail by \cite{2017MNRAS.465.1711B} and references therein. In this method,
we make a map of the $\chi^2$ of the ToA residuals
as a function of $\cos i$ (which has a constant
probability for randomly oriented orbits) and $h_3$ in the case of 
PSR~J1949+3106 (in order to save computational time from being allocated
to regions with very bad values of $\chi^2$), and $M_c$ in the case of PSR~J1950+2414.
We then transform these $\chi^2$ maps into
a 2-D probability density function (pdf) in the $\cos i$ - $M_c$ plane,
we also translate it into a similar pdf in
the $M_p$ - $M_c$ plane. The contours holding 68.23 and 95.44 \% of all probability 
in these planes are displayed in Figs.~\ref{figure:mass_mass_J1949+3106} for PSR~J1949+3106
and \ref{figure:mass_mass_J1950+2414} for PSR~J1950+2414. For the former, the
region of high probability is well described by the orthometric Shapiro delay
parameters, $h_3$ and $\varsigma$, and their uncertainties, as predicted
by \cite{2010MNRAS.409..199F} and already observed in this system
by \cite{2012ApJ...757...89D}. For the latter system, the region of high probability
is well described by $h_3$, $\dot{\omega}$ and their uncertainties.
Projecting these 2-D pdfs onto the different axes, we obtain the probabilities
for the masses and orbital inclination. For PSR~J1949+3106 the 68.3 \% confidence
limits are: 
$M_c \, = \, 0.81_{-0.05}^{+0.06} \, M_{\odot}$,
$M_p \, = \, 1.34_{-0.15}^{+0.17} \, M_{\odot}$ and
$i \, = 79.9_{-0.9}^{+0.8}\, \deg$; for PSR J1950+2414
$M_c \, = \, 0.2795_{-0.0038}^{+0.0046} \, M_{\odot}$,
$M_p \, = \, 1.496\, \pm \, 0.023 \, M_{\odot}$ and
$i \, = 75.7_{-2.8}^{+2.2}\, \deg$. For both systems, these values and
uncertainties are in good agreement with those obtained using
the DDGR model in {\sc tempo}.

\section{Discussion}
\label{sec:discussion}

\subsection{PSR J1949+3106}

The companion WD to this pulsar has a mass of about 0.8 $M_{\odot}$;
this suggests it is a Carbon-Oxygen WD. 
The pulsar in this system seems to have a rather normal mass
($M_p \, = \, 1.34\, M_{\odot}$), although in this case the
precision of the measurement is still consistent, within
2 $\sigma$, with a wide range of masses, from 1.06 to 1.70 M$_{\odot}$.
This means that improvements on the mass measurements are still of
scientific interest.
Since the measurement precision for $\dot{\omega}$
improves faster than for the Shapiro delay parameters
(the uncertainties decrease as $T^{-3/2}$ for the former versus $T^{-1/2}$
for the latter, where $T$ is the timing baseline), it is
likely that, in a not-too-distant future, the combination
of $\dot{\omega}$ with the Shapiro delay parameters will
yield much more precise masses, as in the case of the other
pulsar described in this work, PSR~J1950+2414.
If we set the uncertainty of $M$ to zero in the DDGR model,
we can simulate the results we would obtain for this system if we measured
a very precise $\dot{\omega}$. In that case, the individual masses
can be measured with an uncertainty of only $\sim \, 7.5 \, \times \, 10^{-4} \, M_{\odot}$
from the combination of the ``fixed'' $\dot{\omega}$ and $h_3$.

The characteristics of this system suggest a relatively
mild evolutionary history. The small transverse peculiar velocity
indicates that the kick associated with the
supernova that formed this neutron star was unusually 
small, even though its effect on the peculiar motion of the system would have been
attenuated by the large mass of the progenitor of the WD companion.
This suggests that the envelope of the progenitor to PSR~J1949+3106
was heavily stripped by the progenitor of the companion WD.

In double neutron star systems, there seems to be a positive correlation
between the kick magnitude and the mass of the second-formed NS
\citep{2017ApJ...846..170T}. If kick magnitude is also correlated with 
the mass for the first-formed NS in NS-WD systems, then the
small velocity of PSR~J1949+3106 relative to the LSR and the small implied kick
would suggest a relatively small mass for PSR J1949+3106.

\subsection{PSR J1950+2414}

As discussed already by \cite{2015ApJ...806..140K} and in
section~\ref{sec:introduction}, the second object studied in this work,
PSR~J1950+2414 is a member of a recent and growing class of MSPs
with unexplained large ($e \, \sim \, 0.1$) orbital eccentricities
and orbital periods between 22 and 32 days.

Any measurements of the masses and proper motions for these intriguing systems,
such as those we have obtained above, are important for testing the
hypotheses that have been advanced for their formation.
Apart from PSR~J1950+2414, measurements have just been published for PSR~J1946+3417
\citep{2017MNRAS.465.1711B} and for PSR~J2234+0611 \citep{2019ApJ...870...74S}.
As discussed in detail in these papers,
these mass measurements exclude the hypotheses proposed
by \cite{2014MNRAS.438L..86F} and \cite{2015ApJ...807...41J},
which are based on sudden mass loss of the MSP progenitor.
All measurements thus far are consistent with the expectations of the
hypothesis proposed by \cite{2014ApJ...797L..24A}. This
proposes that the orbital eccentricity is caused by
material ejected from the companion WD by nuclear reactions
happening near its surface. This hypothesis predicts, among other
things, that the MSPs in these systems should have a range of masses
similar to those of the general MSP population
\citep{2016ARA&A..54..401O,2016arXiv160501665A},
which appears to be true.

Regarding the companions to the MSPs in these systems, the prediction
of all hypotheses advanced to date is that they should be Helium
white dwarfs with masses given by the \cite{1999A&A...350..928T} relation. The companion
masses measured for PSR~J1950+2414 (this work) and PSR~J2234+0611
\citep{2019ApJ...870...74S} are in agreement with the
 \cite{1999A&A...350..928T} relation. Furthermore, optical observations
of the companion of PSR~J2234+0611 have confirmed that it is a 
He WD \citep{2016ApJ...830...36A}.
However, the companion mass measured for PSR~J1946+3417
($M_c \, = \, 0.2556(19)\, M_{\odot}$, \citealt{2017MNRAS.465.1711B})
is slightly smaller than that expectation.

One aspect that has not been discussed in detail until now has
been the position of the eccentric MSPs in the $P$-$\dot{P}$
diagram. If, as predicted by \cite{2014ApJ...797L..24A}, these systems
formed essentially as all other MSP binaries, they should be located
in the same areas of the $P$-$\dot{P}$ diagram. This also
appears to be the case as well.

\section*{Acknowledgements}
This work is supported by National Key R\&D Program of China No. 2017YFA0402600.
The Arecibo Observatory is a facility of the National Science Foundation operated under cooperative 
agreement by the University of Central Florida in alliance with Yang Enterprises, Inc. and Universidad 
Metropolitana (NSF; AST-1744119).
The National Radio Astronomy Observatory is a facility of the NSF operated under
cooperative agreement by Associated Universities, Inc.
Pulsar research at Jodrell Bank and
access to the Lovell Telescope is supported by a Consolidated
Grant from the UK's Science and Technology Facilities
Council.
P.C.C.F. and J.W.T.H. gratefully acknowledge financial support by the European Research Council,
under the European Union's Seventh Framework Programme (FP/2007-2013) grant agreements
279702 (BEACON) and 337062 (DRAGNET) respectively,
P.C.C.F. further acknowledges support from the Max Planck Society;
J.W.T.H. from a NWO Vidi
fellowship; J.S.D was supported by the NASA Fermi program; and
W.W.Z. by  the 
Chinese Academy of Science Pioneer Hundred Talents Program, the Strategic Priority Research Program of the Chinese Academy of Sciences Grant No. XDB23000000, and by the National Natural Science Foundation of China under grant No. 11690024, 11743002, 11873067.
K.S., S.C. and J.M.C. are (partially) supported by the NANOGrav Physics
Frontiers Center (NSF award 1430284).
V.M.K. acknowledges NSERC Discovery Grants and the Canadian Institute for Advanced Research (CIFAR)
and further support from NSERC's Herzberg Award, the
Canada Research Chairs Program, and the Lorne Trottier Chair in Astrophysics and Cosmology.
S.M.R. is a CIFAR Senior Fellow.
We thank Nobert Wex and Thomas Tauris for the stimulating discussions and useful suggestions.

\facility{Arecibo}
\software{PRESTO \citep{2011ascl.soft07017R}, PSRCHIVE \citep{2004PASA...21..302H}, TEMPO \citep{2015ascl.soft09002N}}



\end{document}